\documentclass[useAMS,usenatbib]{mn2e}
\usepackage{url,times,graphicx,amsmath,amsfonts,amssymb,aas_macros,color,epsfig,epstopdf,caption,subcaption}
 \usepackage{float}
\usepackage[caption = false]{subfig}

\newcommand{\hMpc}{{\ifmmode{h^{-1}{\rm Mpc}}\else{$h^{-1}$Mpc}\fi}}
\newcommand{\hkpc}{{\ifmmode{h^{-1}{\rm kpc}}\else{$h^{-1}$kpc}\fi}}
\newcommand{\hMsun}{{\ifmmode{h^{-1}{\rm {M_{\odot}}}}\else{$h^{-1}{\rm{M_{\odot}}}$}\fi}}
\newcommand{\ltsima}{$\; \buildrel < \over \sim \;$}
\newcommand{\gtsima}{$\; \buildrel > \over \sim \;$}
\newcommand{\lsim}{\lower.5ex\hbox{\ltsima}}
\newcommand{\gsim}{\lower.5ex\hbox{\gtsima}}

\def\lesssim{\mathrel{\hbox{\rlap{\hbox{\lower4pt\hbox{$\sim$}}}\hbox{$<$}}}}
\def\gtrsim{\mathrel{\hbox{\rlap{\hbox{\lower4pt\hbox{$\sim$}}}\hbox{$>$}}}}

\newcommand{\beq}{\begin{equation}}
\newcommand{\eeq}{\end{equation}}
\def\beqa{\begin{eqnarray}}
\def\eeqa{\end{eqnarray}}
\def\hMpc{$h^{-1}\,{\rm Mpc}$}
\def\hkpc{$h^{-1}\,{\rm kpc}$}

\def\head{
 \vbox to 0pt{\vss
                   \hbox to 0pt{\hskip 440pt\rm LA-UR-10-07069\hss}
                  \vskip 25pt}}

\title[Protoclusters of galaxies in MUSIC]
{The MUSIC of Galaxy Clusters III: Properties, evolution and Y-M scaling relation of protoclusters of galaxies}
\author[F. Sembolini et. al]
       {Federico Sembolini$^{1,2}$\thanks{E-mail: federico.sembolini@uam.es},  Marco De Petris$^2$, Gustavo Yepes$^{1}$, Emma Foschi $^{2}$, Luca Lamagna$^2$\newauthor Stefan Gottl\"ober$^3$\\
$^{1}$Departamento de F\'isica Te\'orica, M\'odulo C-15, Facultad de Ciencias, Universidad Aut\'onoma de Madrid, 28049 Cantoblanco, Madrid, Spain\\
$^2$Dipartimento di Fisica, Sapienza Universit\`a di Roma, Piazzale Aldo Moro 5, 00185 Roma, Italy\\
$^3$Leibniz-Institut f\"ur Astrophysik, An der Sternwarte 16, 14482 Potsdam, Germany
}

\begin{document}

\date{Accepted XXXX . Received XXXX; in original form XXXX}

\pagerange{\pageref{firstpage}--\pageref{lastpage}} \pubyear{2010}

\maketitle

\label{firstpage}

\begin{abstract}
In this work we study the properties of protoclusters of galaxies by employing the MUSIC set of hydrodynamical simulations, featuring a mass-limited sample of 282 resimulated clusters with available merger trees up to high redshift, and we trace the cluster formation back to $z$ = 1.5, 2.3 and 4.
We study the features and redshift evolution of the mass and the spatial distribution for all the cluster progenitors and for the protoclusters, which we define as the most massive progenitors of the clusters identified at $z$ = 0. A natural extension to redshifts
larger than 1 is applied to the estimate of the baryon content also in terms of 
gas and stars budgets: no remarkable variations with redshift are discovered.
Furthermore, motivated by the proven potential of Sunyaev-Zel'dovich surveys to blindly search for faint distant objects, we focus on the scaling relation between total object mass and integrated
Compton $y$-parameter, and we check for the possibility to extend the mass-observable paradigm to the protocluster regime, far beyond the redshift of 1, to account for the properties of the simulated objects.
We find that the slope of this scaling law is steeper than what expected for a self-similarity assumption among these objects, and it increases with redshift mainly for the synthetic clusters where radiative processes, such as radiative cooling, heating processes of the gas due to UV background, star formation and supernovae feedback, are included.
We use three different criteria to account for the dynamical state of the protoclusters, and find no significant dependence of the scaling parameters from the level of relaxation. Based on this, we exclude that the dynamical state is the cause of the observed deviations from self-similarity.

\end{abstract}
\noindent
\begin{keywords}
  methods: numerical -- galaxies: clusters -- 
  cosmology -- cosmology: theory -- cosmology : miscellaneous
\end{keywords}

\section{Introduction} \label{sec:introduction}

The formation of today's large scale structures, from massive clusters
to smaller groups of galaxies, starts from high redshift overdensities
lying along the dark matter filamentary structure known as the cosmic web. 
In the early phases of their evolution these objects are characterized 
by relatively smooth peaks in the spatial distribution of dark matter 
and galaxies, and grow into denser and larger concentrations of dark matter, 
gas, and galaxies at later epochs. Therefore, by systematically searching 
for protoclusters, and studying their dynamics, evolution and abundance
as a function of mass and redshift, it is possible to explore the 
high-z stages of the assembly of present day clusters, and possibly to shed
light on the processes which affect the growth of structures on the tail of 
the halo mass function just as they shape from small overdensities, on the 
verge of virialization, into the largest, most massive bound objects 
in the Universe. The evolution of the halo MF puts constrain on $\Omega_m$
 mass and redshift up to the very early stages of their assembly.

Currently, many observational and theoretical issues impose
critical limitations to this kind of studies:
distance limits the quantity and accuracy of available observations
of these objects. Several direct and indirect approaches have
been tried to perform systematic searches of protoclusters in the
high-z universe, but none of them has been proved to be generally
successful and therefore none has been employed for systematic
protocluster searches up to now. (see sec. 2 for a review of current
observing methods).
A reliable observational proxy for their total mass, which assumes
an insight of the structure assembly at high redshift and a
proper validation of available proxies, as with low-z clusters and
groups, due to the necessity of exploring the mass distribution of
protoclusters
Following the same approach of cluster studies performed up
to redshift 1 \citep{IO}, in this work we explore the
possibility to observe protoclusters through the detection of their
thermal Sunyaev-Zeldovich (th-SZ, \citealt{SZ70})
imprint into the Cosmic Microwave Background (CMB). Due to
the lack of dimming of the scattered CMB photons off ionized gas
in the high-z halos, and to the uniqueness of its spectral signature
at mm/submm wavelengths, th-SZ effect appears as a viable tool
for high-z object-finding, as proved from the success of blind cluster surveys 
from the current generation of millimeter telescopes 
(\citealt{STA09}, \citealt{MA11}, \citealt{PLANCK11}, \citealt{WI11}, 
\citealt{PLANCK29}, \citealt{RE13}).
In principle, the high angular resolution and the sensitivity needed 
to provide reliable SZ detections of farther, fainter, less evolved objects 
should be at hand, thanks to the upcoming generation of instruments 
like Mustang2, ALMA, CCAT, SPT3G (among others) or satellite missions 
like Millimetron\footnote{{\ttfamily http:// www.sron.rug.nl/millimetron}} or the proposed PRISM 
\footnote{{\ttfamily http://www.prism-mission.org}} \citep{PRISM}. 

Within a self-similar scenario of structure formation, a tight
correlation between an aperture-integrated th-SZ signal 
(which is a measure of the total thermal energy of the hot gas
in a large virialized structure) and the total mass M of the object,
is expected. For clusters and groups, this scaling law, and its small
deviations from self-similarity, have been studied through semi-analytical approaches (e.g. \citealt{SH08}, \citealt{SUN11})
simulations of cosmological volumes (\citealt{BATTAGLIA11}, \citealt{KAY12}, 
\citealt{IO}) and verified through observations (\citealt{BONA08}, 
\citealt{MAR12}, \citealt{PLANCK3}, 
\citealt{PLANCK20} , \citealt{SIF13}). 
In this paper we verify for the first time the extension of the
self-similarity assumption to the progenitors of today�s clusters. For
this purpose we use synthetic objects extracted from a large dataset
of hydrodynamical simulations of clusters of galaxies: MUSIC.
While the definition of a protocluster is in debate from an observational
point of view (details are provided in sec. 3), the availability
of numerically simulated structures at all the ages up to $z$ = 4 may
allow to trace the evolution of clusters back to the formation
through the information of the merging tree for each object.
The paper is organized as follows. In Section 2 observational
approaches of distant galaxies, assumed as possible progenitors of
groups and clusters, are reported. The protoclusters extracted from
simulation are described in Section 3, where also the baryons budget
is explored in terms of gas and star fractions for different masses
and protocluster redshifts. Considerations about the spatial distribution
of the progenitors and useful criteria to quantify the virial
state of these objects are also treated. The validity of the self similarity
approach, basics for the Y - M scaling law, is verified in
Section 4 for objects ranging from z = 1 up to z = 4. In Section 5
we summarize and discuss our main results.

\section{Multiwavelength Observations of protoclusters of galaxies} 

In order to investigate when and how clusters are formed, it is necessary to obtain a sample of objects at $z > 1$. In the past decade, there has been a significant increase in the study of clusters at redshifts up to 1, while the difficulty of observing protoclusters of galaxies limits the amount and accuracy of the observations and surveys that are available.

In fact, notwithstanding the development of a new generation of telescopes, many observational and theoretical issues impose critical limitations to these kinds of studies. The hindrances in observing protoclusters of galaxies are linked to the relative low angular resolution of the observation instruments used and consequently to the inability to investigate extended structures.  Moreover, according to the $\Lambda$CDM model, it is extremely rare to find objects with $M > 3\times 10^{14}$ $M_{\odot}$ at $z  > 1$ \citep{SP05} and high redshift galaxies do not dominate the number counts in surveys. In addition, X-ray emission becomes too faint to be measured since the surface brightness decreases as $(1 + z)^4$ . Despite these limitations, observations made in the optical/IR wavelengths together with the XMM-Newton have identified an overdensity of galaxies emitting at $z = 1.579$ in the X-rays band \citep{JS11}. Another finding in the survey XDCP (XMM-Newton Distant Cluster P
 roject) led to the identification of a low-mass (proto)cluster ($M = 10^{14}$ $M_{\odot}$) at $z = 1.1$, thanks to the multi-band observation with the GROUND imager \citep{PI12}. X-rays observations thus make possible to observe groups and clusters at much earlier stages (i.e., $z \ge 1$). However, these surveys still remain subject to the selection effects. 

The ability to perform systematic searches of protoclusters in the high-$z$ universe has long been sought after.  Various methods have been applied in order to render this possible. High redshift galaxies can be distinguished from the profuse nearby galaxies due to some peculiar spectral characteristics. Thanks to these features, it is possible to use other methods of detection in order to investigate the universe at high redshift.
One of the methods most widely used is targeting high-$z$ radio galaxies (HzRGs). These are massive star forming galaxies with enormous radio luminosities (\citealt{MDB08}, \citealt{SEY07}, \citealt{RV04}). According to the model of hierarchical galaxy formation, it is possible to find galaxy overdensities around HzRGs (\citealt{ST03}, \citealt{MAY12}), which should be likely surrounded by cluster progenitors (\citealt{VE07}, \citealt{KU10}, \citealt{HAT10}, \citealt{WI13}). 

Recent results were obtained as part of the HeRGE project. A study of the IR spectral energy distribution of the Spiderweb Galaxy at $z = 2.156$ showed that this protogalaxy is in a particular phase implying both of AGNs and starburst \citep{SEY12}. Moreover, by combining different studies of the environment of this galaxy, it was possible to identify several protocluster members surrounding the host galaxy, with an estimated mass $> 2 \times 10^{14} M_{\odot}$ within a region of 3 Mpc.
\begin{figure}
  \centering  \includegraphics[angle=0,width=8cm]{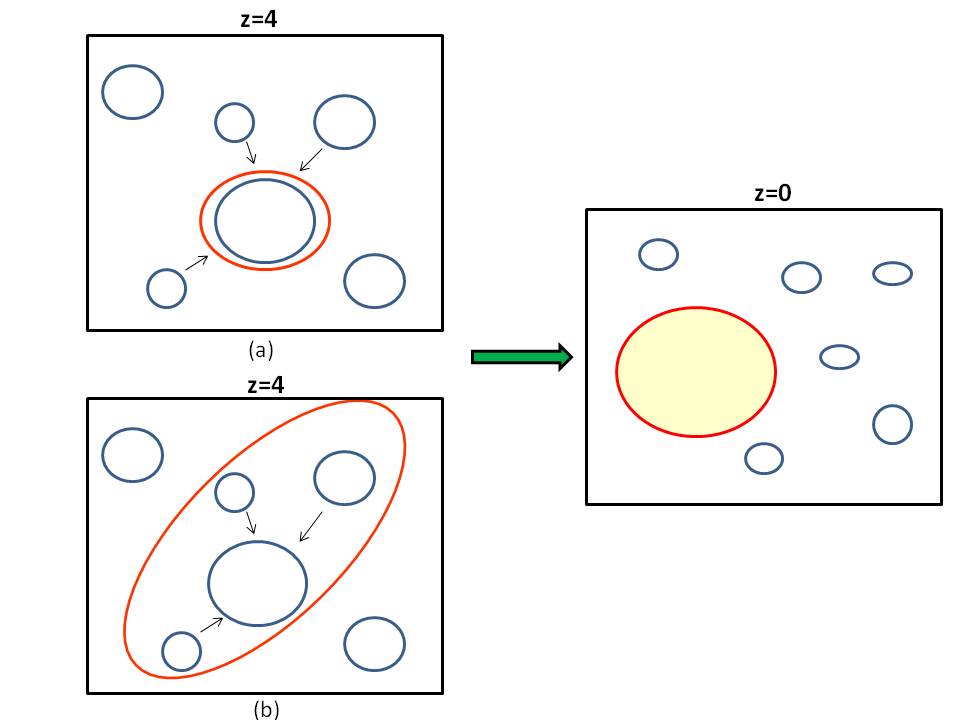}
  \caption{Schematic representation of the two protocluster definitions. In the panels on the left, the red circle confines the protocluster at $z$ = 4 according to the first definition (a) and the second (b). In the panel on the right, it is shown the representation of the present-day cluster, which is the main object at $z$ = 0, formed during the evolution process of the protocluster (according to both definitions).}
  \label{proto_def}
\end{figure}
\begin{figure*}
\centering\includegraphics[angle=0,width=18cm]{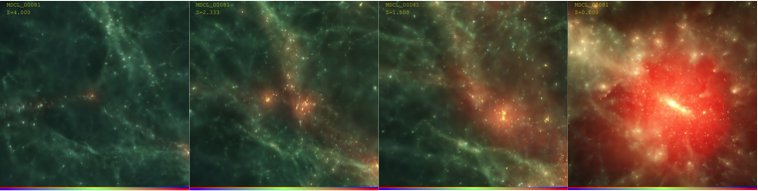}
\caption{The evolution of one of the resimulated regions of MUSIC-2 dataset (CSF run) from high redshift to$z$ = 0 (from left to right: $z$ = 4, 2.3, 1.5, 0). At high redshifts the protocluster is located in the top part of the image. All the images of MUSIC clusters have been generated with SPLOTCH \citep{SPLOTCH} and are available at the website {\ttfamily http://music.ft.uam.es}.}
\label{cl2}
\end{figure*}

Another approach is based on the selection of the Lyman Break Galaxies (LBGs), which are star-forming galaxies at $2.5 < z < 5$ characterized by the Lyman break at 912 $\AA$ in the rest-frame \citep{GIA02}. Searching for Ly$\alpha$ emitters is another way to identify galaxy cluster progenitors \citep{ST98}. Star-forming galaxies exhibit strong emissions of this particular line because Ly$\alpha$ photons are resonantly scattered in neutral hydrogen. Using narrow-band imaging, it is possible to search for overdensities of line emitting objects at a specific redshift. Many studies have been conducted using this particular method and have resulted in the discovery of cluster progenitors beyond $z = 3$ (\citealt{MAT09}, \citealt{ST00}, \citealt{YA12}, \citealt{CAP11}).

It is also possible to observe star-forming galaxies by detecting their sub-millimeter emission. In fact, the large and negative $k$-correction \citep{BL93}, due to the steepness of the submm-spectra, makes the high-redshift galaxies more detectable than their low-redshift counterparts. It is possible to identify star-forming galaxy at FIR/submm wavelengths, through the detection of dust emission. Studies of radio galaxies have been concentrated on high redshift objects since their submm luminosity increases with redshift and their emission is in correspondence with the peak of dust emission \citep{ARC01}. A population of almost 200 luminous galaxies at $z > 1$ has been revealed through deep surveys in the submm/mm waveband thanks to detectors such as SCUBA, MAMBO and BOLOCAM \citep{BL02}.

Thanks to the current ground-based projects (such as ACT, SPT), it is becoming increasingly possible to observe high-z objects while steadily increasing the redshift through the SZ effect.
Recently, it has been made possible to detect clusters at redshift greater than 1 via the Sunyaev-Zel'dovich effect.  The SPT-SZ survey allowed the identification of the highest redshift galaxy cluster that was seen via the SZ effect, which is at $z = 1.478$ \citep{BAY13}. 


\section{Protoclusters of galaxies in the MUSIC dataset}\label{sec:bar}
Since it is difficult to prove how and when today's clusters of galaxies were formed, what is meant by the term ''protocluster'' from an observational point of view is in debate. 
As a result, it is particularly important to be able to discriminate in this study all the high-$z$ objects related to present clusters. With this purpose we define as progenitors all those objects which will merge during the cluster evolution to form and be part, with at least a consistent fraction of their mass (see Sec.\ref{MR}), of the cluster observed at $z$ = 0.

For the purpose of our work, two alternative and general definitions have been used. We assume as a protocluster (Fig.\ref{proto_def}):
\begin{enumerate}
\item  the most massive halo at high redshift among all the progenitors;
\item the ensemble of all the progenitors with a mass larger than a selected value (which depends on limits on the observability or on the resolution of the simulation)
\end{enumerate}
According to this, numerical simulations constitute the ideal tool to define  and study protoclusters: in fact, using a merger tree, it is straight forward to trace back at high redshift the particles, and therefore the progenitors, which will end up into a virialized cluster at $z$ = 0. This fundamental characteristic allows to overcome the principal problem found in observations, where it is impossible at present day to be completely confident if a massive object observed at high redshift will actually evolve into a cluster during its history.

To the scope of this work, which is to study some integrated properties of protoclusters, we choose to adopt the first definition of the two aforementioned. Most of the analysis shown in this work is referred to protoclusters, though it is also interesting to make some considerations about all the progenitors in terms of their mass and spatial distributions.

\subsection{The simulations}
The simulations used in this work are part of the MUSIC dataset\footnote{Initial conditions and snapshots of MUSIC clusters, plus many images, are publicly available at the webpage  {\ttfamily http://music.ft.uam.es}}. A detailed description of the MUSIC dataset can be found in \cite{IO}, so in this subsection we will limit to recall some main characteristics of the simulations and to define the subset that we selected for our analysis. The protoclusters presented in this work have been taken from the MUSIC-2 dataset, a mass selected volume limited sample of resimulated clusters extracted from the MultiDark Simulation (MD, \citealt{PETER12}). From the MD simulation, a dark-matter only simulation of 2048$^3$ particles in a (1$h^{-1}$Gpc)$^3$ cube, all the objects with a total virial mass $M_{vir} > $10$^{15}$\hMsun at $z$ = 0 in the low resolution version of the simulation were selected and resimulated adding SPH and star particles, plus various radiative processes (including r
 adiative cooling, heating processes of the gas
arisen from a UV background, star formation and supernovae feedback). 

In total, 282 lagrangian regions with a radius of 6$h^{-1}$Mpc surrounding a massive cluster were resimulated. All clusters were resimulated, with the same zooming techniques and resolution, both with radiative (CSF subset, see Fig.\ref{cl2}) and non-radiative physics (NR subset). The mass resolution for these simulations corresponds
to $m_{DM}$=9.01$\times$10$^8$\hMsun and to $m_{gas}$=1.9$\times$10$^8$\hMsun. The parallel TreePM+SPH GADGET code (\citealt{GAD05})
was used to run all the resimulations. Among the 15 snapshots describing the evolution of each cluster in the redshift range 0 $\leq z \leq$ 9, we concentrate on those corresponding to $z$ = 1.5, 2.3, 4.0, assuming that at $z \leq$ 1 all objects have already evolved into clusters. The analysis shown hereafter is therefore focused on the protoclusters corresponding to the most massive progenitors of the most massive clusters of each of the 282 MUSIC-2 resimulated regions. Among all the massive clusters at $z$ = 0, almost 50 per cent have $M_{vir} >$ 10$^{15}$\hMsun and almost all $M_{vir} >$ 5$\times$ 10$^{14}$\hMsun.

\begin{figure}
\centering\includegraphics[angle=0,width=8cm]{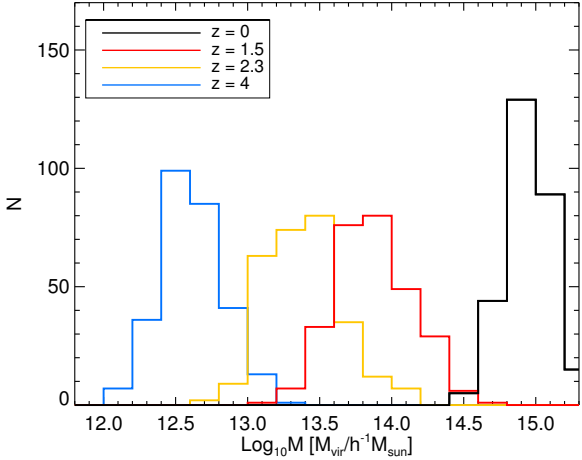}
\caption{Distribution of virial mass of protoclusters at the different analyzed redshifts ($z$ = 1.5 in red, $z$ = 2.3 in yellow, $z$=4 in blue), compared with the mass distribution of the same objects evolved into clusters at $z$ = 0 (black).}
\label{mdist}
\end{figure}

\begin{figure}
\centering\includegraphics[angle=0,width=8cm]{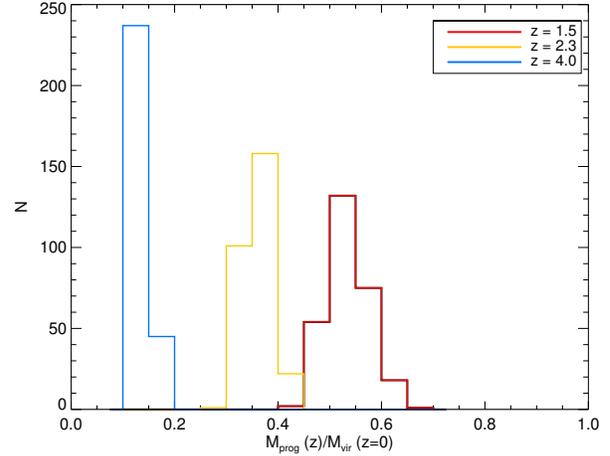}
\caption{Distribution of the mass fraction of the total cluster mass at $z$ = 0 hosted by progenitors with M$>$1.2$\times 10^{10} h^{-1}M_{\odot}$ at $z$ = 1.5 (red), 2.3 (yellow) and 4 (blue).}
\label{Castor}
\end{figure}

\subsection{Mass and spatial distributions of progenitors and protoclusters}\label{MR}
As aforementioned, we can use a merger tree to track back in time the cluster history and individuate all the progenitors (including the protocluster) at high redshifts. We use the merger tree of the Amiga Halo Finder (AHF, \citealt{KK09}) to select all the high redshift objects containing particles which will be part of a massive cluster at $z$ = 0, and, according to the definition given at the beginning of this section, we individuate as progenitors all those halos whose at least the 80 per cent of their particles are found to be part of the cluster formed at $z$ = 0. Considering that AHF is able to discern all halos constituted by at least 20 particles, we can list all progenitors with $M >$ 1.2$\times$10$^{10}$\hMsun.

It is interesting to study the mass distribution of protoclusters (calculated at the virial radius) to explore the mass evolution with redshift and to compare the mass of protoclusters with that of the other progenitors, in order to check at each redshift whether the protoclusters show already a mass sensitively bigger than the other progenitors. For each halo the virial radius $R_{vir}$ is computed, defined as the radius at which the mean internal density is $\Delta_{vir}$ times the background density of the Universe at that redshift (the value of $\Delta_{vir}$ therefore depends on redshift too). According to this, the definition of virial mass is\footnote{$\rho_c $ is the critical density of the Universe, defined as $\rho_c$=3$H^2$/8$\pi G$ ($H$ is the Hubble constant and $G$ the gravitational constant)} :
\beq
M_{vir} = \frac{4\pi}{3}\Delta_{vir}\Omega_m\rho_{crit}R_{vir}^3
\eeq
In Fig.\ref{mdist} the distribution of $M_{vir}$, at the 3 considered redshifts, is
compared with the mass distribution of the evolved clusters at $z$ =
0. 
\begin{figure}
\centering\includegraphics[angle=0,width=8cm]{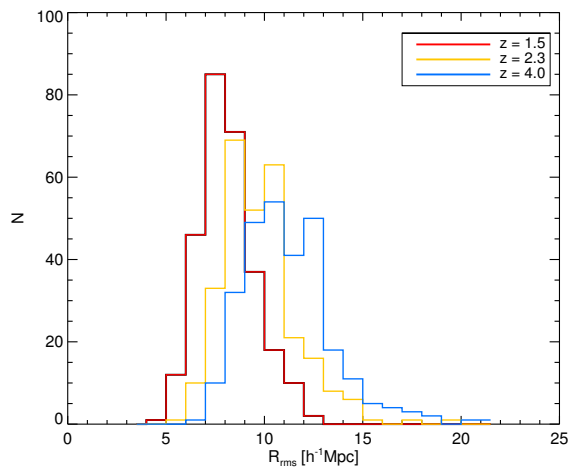}
\caption{Spatial distribution of progenitors in terms of $rms$ distances at $z$=1.5 (red), 2.3 (yellow) and 4 (blue), all calculated assuming the center of mass of the protocluster as the center of the cluster formation region.}
\label{Castor}
\end{figure}
The mass distributions of protoclusters are almost completely separated among each other at the 3 different redshift considered (though showing a large dispersion, see Tab.1): at $z$ = 4 most of halos have a mass of a few times 10$^{12}$\hMsun, with only a small number of objects with $M >$ 10$^{13}$\hMsun; at $z$ = 2.3 almost all objects show masses in the range 10$^{13} < M[$\hMsun $]<$ 10$^{14}$; at $z$ = 1.5 we find more than 100 halos with $M >$ 10$^{14}$\hMsun and that therefore can be also considered as already evolved into clusters (if we define as cluster a virialized halo with $M >$ 10$^{14}$\hMsun). 

The mean value of each mass distribution is reported in Tab.1. It is interesting to measure the fraction of the total mass of the cluster at $z$ = 0 which is contained in the progenitors at high redshift, $M_{prog}/M(z=0)$: we find that at $z$ = 4 only a very small fraction of the total mass (14 per cent, see Tab.1) is hosted by the progenitors, showing how at this age of the Universe most of the matter which will collapse into clusters is still in the form of diffuse matter (i.e. filaments) or of structures under galaxy size; still at $z$ = 1.5 only almost half of the total mass of clusters at $z$ = 0 is still not detected in progenitors with $M >$ 1.2$\times$10$^{10}$\hMsun. It is also worth mentioning that mass ratio between the second most massive progenitor and the protocluster itself is in mean about 70 per cent at $z$ = 4 and still almost 60 per cent at $z$ = 1.5, an evidence that during their formation history most of massive clusters go through a major merger at $z >
 1$.

We also concentrate on studying the spatial distribution of progenitors at different redshifts; if we assume the center of the region of the forming cluster as the center-of-mass of the protocluster, we can define the root mean square distance as:
\beq
R_{rms} = \sqrt{\frac{\sum_{i=0}^N r_i^2}{N} }
\eeq
where $N$ is the total number of progenitors and $r_i$ the distance between the $i$-th progenitor and the center-of-mass of the protocluster. The distribution of the $R_{rms}$ at the 3 redshift analyzed is shown in fig.\ref{Castor} and the mean values reported in Tab.1: the initial cluster forming region shows in mean $R_{rms}\sim$ 11$h^{-1}$Mpc at $z$ = 4, contracted at $R_{rms}\sim$ 8$h^{-1}$Mpc at $z$ = 1.5. We remind that the typical virial radius at $z$ = 0 of the clusters formed by these regions collapsing is about 2$h^{-1}$Mpc and the mean virial radii of our dataset at different redshifts are listed in Tab.1. It is interesting to observe that even if we pull down from 80 to 50 or 20 per cent the threshold of particles of an object which have to be part of the cluster at $z$ = 0 in order to define it as a progenitor, the mean values of the $R_{rms}$ do not change of more than 5 per cent; we find that the maximum radius (whose mean values are also reported in Tab.1) of 
 the cluster forming area is always $R_{max}\sim$ 2$R_{rms}$.

\begin{table}
\begin{center}
\begin{tabular}{|c|c|c|c|}
\hline
  & \textbf{z = 1.5} & \textbf{z = 2.3} & \textbf{z = 4.0} \\
\hline
\textbf{M$_{vir}$ [10$^{13}$\hMsun]} & 9.2$\pm$6.4 & 3.0$\pm$2.2 & 0.5$\pm$0.3 \\
\hline
\textbf{R$_{vir}$ [$h^{-1}$Mpc]} & 1.04$\pm$0.22 & 0.72$\pm$0.15 & 0.40$\pm$0.07 \\
\hline
\textbf{M$_{prog}$/M($z=0$)} & 0.53$\pm$0.04 & 0.36$\pm$0.02 & 0.14$\pm$0.01 \\
\hline
\textbf{R$_{rms}$ [$h^{-1}$Mpc]} & 8.1$\pm$1.4 & 9.8$\pm$1.9 & 11.2$\pm$2.3 \\
\hline
\textbf{R$_{max}$ [$h^{-1}$Mpc]} & 16.1$\pm$2.6 & 19.2$\pm$3.1& 22.0$\pm$3.7 \\
\hline
\end{tabular}
\end{center}
\caption{Mean values (at different redshifts) of the virial mass (M$_{vir}$) and virial radius (R$_{vir}$) of protoclusters, of the mass fraction hosted by progenitors with $M >$ 1.2$\times$ 10$^{10}$\hMsun (M$_{prog}$/M($z=0$), of $R_{rms}$ and $R_{max}$.}
\end{table}

\subsection{Baryon properties of protoclusters} \label{sec:theory}
It is interesting to explore the baryon content of protoclusters of galaxies, in order to follow the evolution of the baryon, gas and star fraction (respectively $f_b, f_g, f_s$) of galaxy clusters in the range 0 $\leq z \leq$ 4; at the same time, we can check whether our dataset is affected by cold flows or galaxy feedbacks, effects that usually affect the inner regions of clusters but that in the case of such small objects could affect also areas closer to the virial radius. This is important to verify if we want to study integrated properties of protoclusters, such as the integrate Compton parameter $Y$, directly depending on the gas content, which has therefore to be described correctly (see Section 4).

The baryon, gas and star fractions are defined by simply taking into account all the gas and star particles falling inside the virial radius:
\beq
f_{b,g,s}(<R_{vir}) = \frac{M_{b,g,s}(<R_{vir})}{M(<R_{vir})}
\eeq
\begin{figure}
\centering\includegraphics[angle=0,width=8cm]{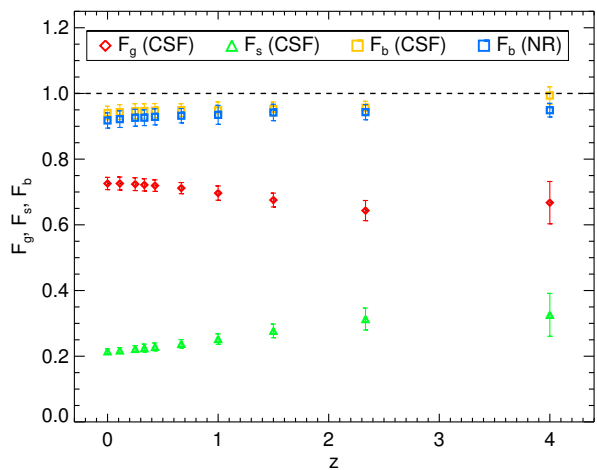}
\caption{Evolution of normalized gas (red diamonds), star (green triangles) and baryon fraction (yellow squares for CSF protoclusters and blue squares for NR) from $z$ = 4 to $z$ = 0.}
\label{fgas}
\end{figure}
where $M_g$ is the mass of the gas, $M_s$ the mass of the star component, and the total baryon mass is defined as $M_b$=$M_g+M_s$. Fig.\ref{fgas} shows the behavior of the mean baryon, gas and star fractions calculated at the virial radius in the redshift range 0 $\leq z \leq$ 4 (results referring to $z \leq$ 1 are taken from \citealt{IO}) , normalized to the critical cosmic ratio $\Omega_b/\Omega_m$, (which according to the cosmology adopted by MUSIC is 0.174) in order to make a comparison with other works adopting different cosmological parameters (since now on we denote the normalized values of $f_{b,g,s}$ using capital letters: $F_{b,g,s}$). The normalized $F_b$ is around 95 per cent at all redshifts for both CSF and NR subsets, as expected slightly higher than what measured in simulations of clusters at $z \leq$ 1 at $\Delta_c$=500, $F_{b,500}\sim$ 0.85 (\citealt{IO}, \citealt{PL13}). We remind that $\Delta_c$ defines that the overdensity is calculated with respect to th
 e mean critical overdensity of the Universe at the redshift analysed. This was easy to predict as the value of the baryon fraction is expected to approach the cosmic ratio going from inner to outer regions of (proto)clusters) and does not vary with redshift. 

The gas fraction appears to be lower for protoclusters ($z \geq$ 1.5, $F_g\sim$ 0.65)
than for clusters ($z \leq$ 1), for which we find that $F_g$ is around 75 per cent, values that can be treated as reasonable if we consider that CSF simulations show $F_g \sim$ 0.65 at  $\Delta_c$=500. The mean value of the normalized star fraction rises from $F_s\sim$ 0.2 at $z$ = 0 to $F_s\sim$ 0.3 at $z$ = 4, an amount still smaller than what is estimated in the inner regions of clusters (those which are more likely to be affected by cold flows) at $\Delta_c$=2500. These results are comforting to the purpose of our analysis, as they allow us to state that there are no dramatic differences in the baryon content between clusters and protoclusters, and we can go on studying the integrated properties depending on gas content of the second ones.

\subsection{Dynamical state of protoclusters}
It is interesting to study the dynamical state of protoclusters in order to check if the morphology of these objects could have an impact on scaling relations.
Three different criteria  are  commonly used to define the morphological state of clusters and protoclusters, aiming at distinguishing relaxed objects from disturbed ones (\citealt{SH06}, \citealt{TUKA}):
\begin{itemize}
\item The presence of mergers, defining as major mergers those objects with a mass bigger than one half of the main object and as minor mergers those objects with a mass between 0.1 and 0.5 times the mass of the main object. Clusters experiencing or having experienced merger processes are more likely to be morphologically disturbed.
\item The center-of-mass offset, namely the spatial separation between center-of-mass of the protocluster and the  center-of-density (maximum density peak), normalized to the virial radius (see eq.\ref{eqr}). Objects showing an high value of $\Delta r$ are considered disturbed.
 \item The accomplishment of the virial theorem, calculating the virial ratio $\eta$ = 2$T$/$\mid$$U$$\mid$ (where $T$ is the kinetic energy and $U$ the potential energy).  If the object is relaxed,  we should find $\eta \sim$1. 
 
 The first criterium seems not to be successful when applied to protoclusters, as these show merger rate much lower than clusters at low redshift (35 per cent for clusters at $z <$ 1). 
 
 We have therefore to concentrate on the two other methods to fulfill our purpose of  distinguishing  relaxed protoclusters from disturbed ones. The center-of-mass offset is quantified ad:
 \beq
 \Delta r = \frac{\mid r_{\delta}-r_{cm}\mid}{R_{vir}}
 \label{eqr}
 \eeq
 where $r_{\delta}$ is the position of the center-of-density of the halo, $r_{cm}$ the center-of-mass and $R_{vir}$ the virial radius. $\Delta r$ is used to quantify substructures
statistics, providing an estimate of the halo's deviations from smoothness and spherical symmetry. 
Up to now, this topic has been treated in literature in the case of dark-matter only simulations. To identify an halo as a relaxed one the limit value, assigned to $\Delta r$, ranges from $\Delta r\leq$ 0.04 \citep{MAC07} to $\Delta r\leq$ 0.1 \citep{DON07}. Here we will show  that hydrodynamical simulations of clusters present higher values of $\Delta r$ with respect to dark-matter only simulations, so we choose to adopt  the highest value among those previously cited: $\Delta r\leq$ 0.1  to define an object as relaxed.

The third and last requirement uses the virial theorem to determine which halos are not dynamically relaxed. The standard definition for a dynamical  system in equilibrium is usually represented by $\eta\sim$1.  Nevertheless,  the effect of those particles situated outside the virial radius but still gravitationally bound to the halo has to be taken into account and included in the estimate of the kinetic and potential energies. These particles are still bound to the halo and their contribution to the virial theorem has to be considered. The additive term, to include this surface pressure energy at the boundary of the halo, can be quantified as \citep{CHANDRA}:
\beq
E_s=\int P_s(r)\boldsymbol{r\cdot dS}
\eeq
The measure of the virial ratio $\eta$ has therefore to be modified to take into account  this  surface pressure term. Therefore, a modified  definition of the virial parameter can  be expressed as follows:
\beq
\eta_1=\frac{2T-E_s}{\mid U\mid}
\eeq
\end{itemize}
Assuming an ideal gas, the surface pressure  can be calculated as \citep{SH06}:
\beq
P_S=\frac{1}{3V}\sum_i(m_iv_i^2)
\eeq
where $V$ is the volume of the spherical shell between 0.8 and 1.0 R$_{vir}$ and $m_i$ and $v_i$ are the mass and velocity of the $i$-th particle respectively. Integrating $P_S$ over the bounding surface of the  halo volume  it is found $E_S$=4$\pi$$r_{med}^3P_S$, assuming $r_{med}\simeq$ 0.9 R$_{vir}$ \citep{TUKA}.\\
We apply this  analysis, already performed by \cite{TUKA} and \cite{POW12} to dark-matter (DM) only halos, to our hydrodynamically simulated protoclusters, in order to check any dependence between halos mass, virial ratio and center-of-mass offset. 

\begin{figure}
\centering\includegraphics[angle=0,width=8cm]{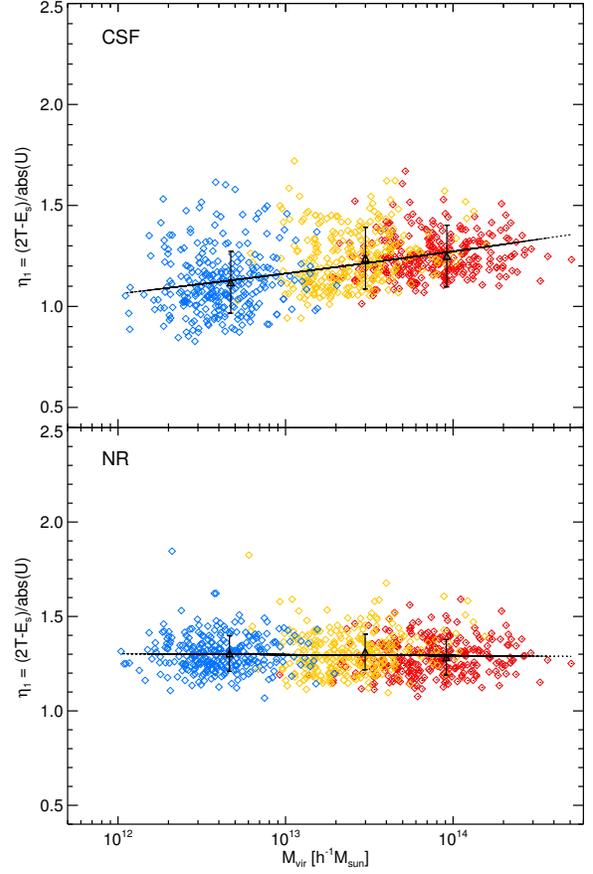}
\caption{Relation between $\eta_1$ and the mass of the progenitors at different redshifts for CSF protoclusters (top panel) and for NR protoclusters (bottom panel): the first one shows a weak but clear dependence of $\eta_1$ with mass, the second one does not. Different colors signs objects at different redshifts ($z$ = 1.5 is red, $z$ = 2.3 is yellow and $z$ = 4 is blue). The black line is the best-fit and the black squares show the mean $\eta_1$ at each redshift.}
\label{etaM}
\end{figure}

In the case of CSF objects, we find  a mild dependence between the mass of the progenitor at different redshifts, $M_z$, and virial ratio $\eta_1$ (see Fig.\ref{etaM}):
\beq
\eta_1 \propto M_z^{0.04}
\eeq
It is interesting to notice how this mass dependence is not fulfilled  by NR clusters  (Fig.\ref{etaM}, bottom panel).  

The numerical values of $E_s$ are generally almost one order of magnitude smaller than the kinetic energy. The correction due to the surface pressure makes therefore the value of the virial ratio lower, but in most cases not enough to reach the expected value of 1.
\begin{figure}
\centering\includegraphics[angle=0,width=8cm]{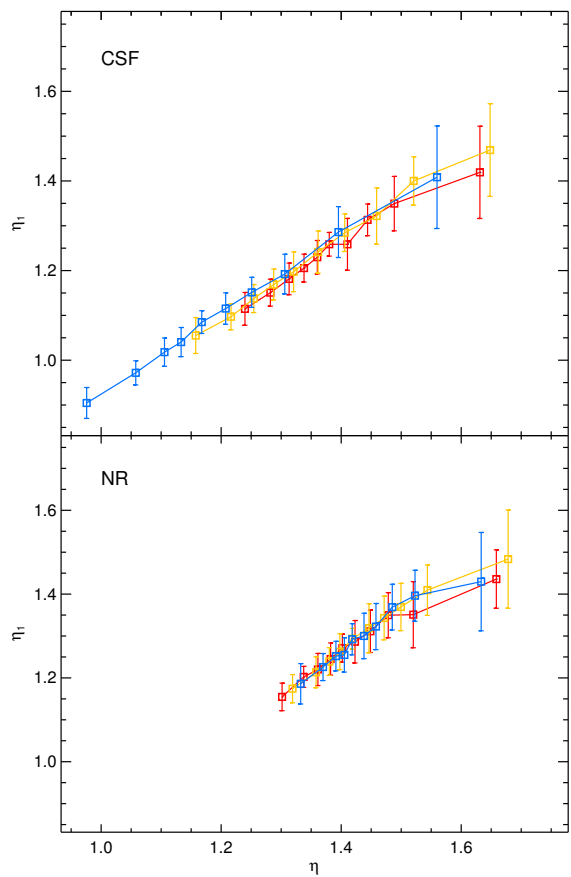}
\caption{$\eta_1-\eta$ relation for CSF protoclusters (top panel)  and NR protoclusters (bottom panel): $z$=1.5 is red, $z$ = 2.3 is yellow and $z$ = 4 is blue. In the CSF relation there is a linear dependence between the two parameters; objects at redshift $z$=1.5 show lower values of $\eta,\eta_1$, reflecting the $\eta_1-M$ linear dependence: objects at $z$=1.5 have bigger masses and therefore higher $\eta_1$ values. NR protoclusters exhibits the same linear dependence than CSF objects, but the values of $\eta$ are distributed on a narrower range at high values.}
\label{etaeta}
\end{figure}

\begin{figure}
\centering\includegraphics[angle=0,width=8cm]{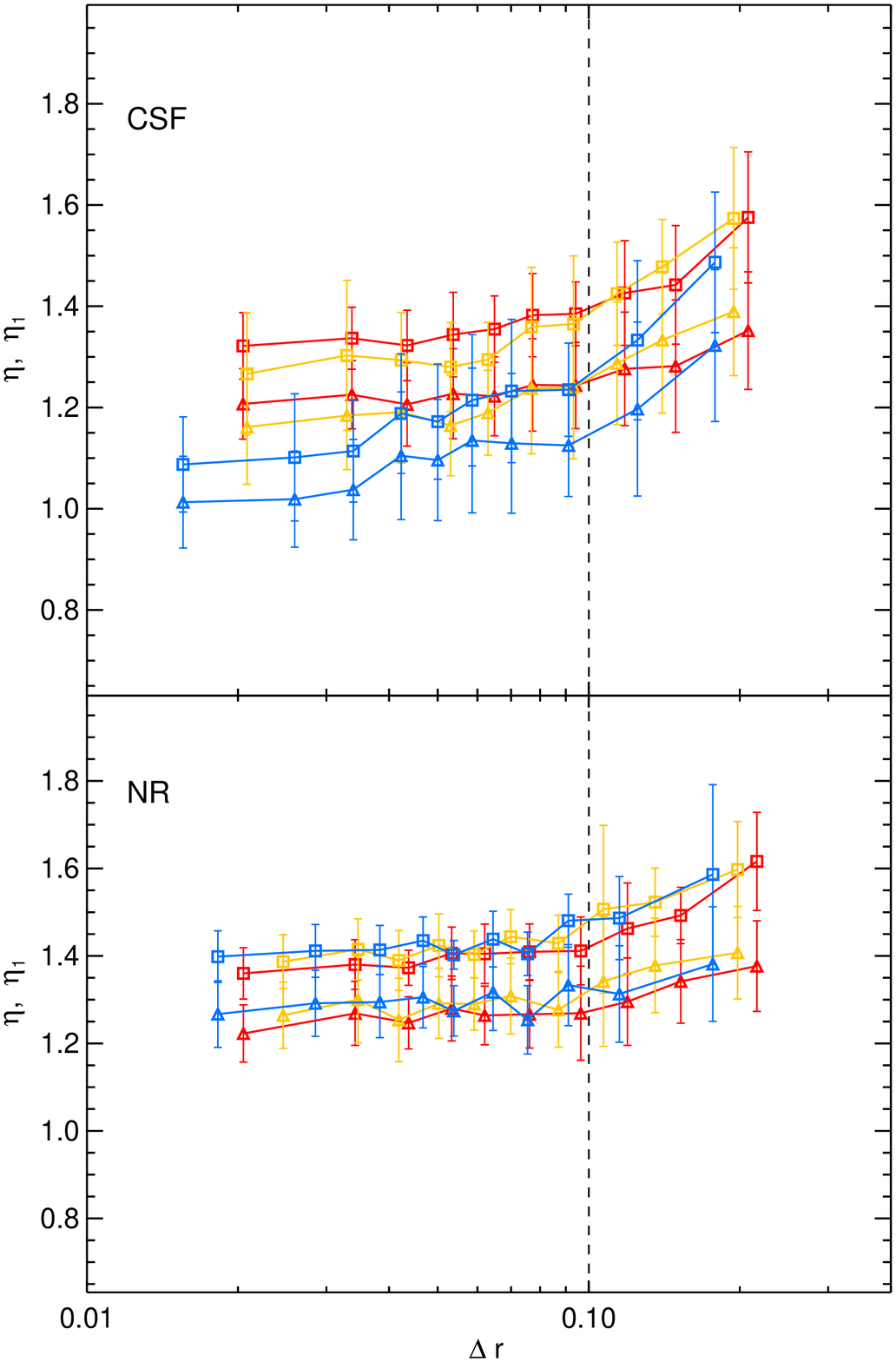}
\caption{Relation between the two definitions of virial ratio $\eta$ (squares) and $\eta_1$ (triangles) and the center-of-mass offset $\Delta r$ at different redshifts ($z$ = 1.5 in red, $z$ = 2.3 in yellow, $z$ = 4 in blue) for the CSF (top panel) and NR (bottom panel) subsets. The vertical black dashed line indicate the upper value to define relaxed halos ($\Delta r$ = 0.1).}
\label{deltar}
\end{figure}

\begin{figure}
\centering\includegraphics[angle=0,width=8cm]{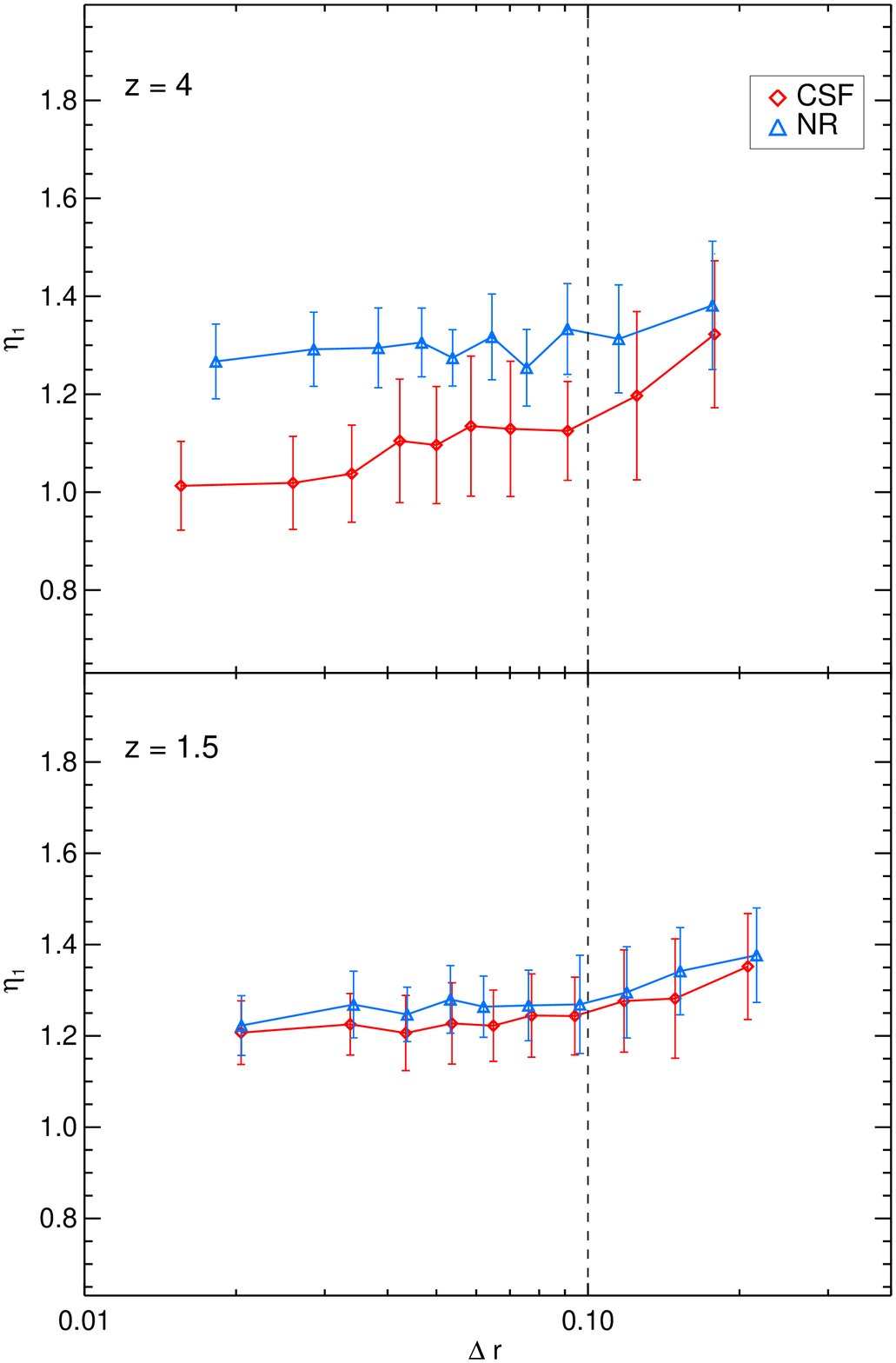}
\caption{$\eta_1$-$\Delta r$ relation at $z$ = 4 (top panel) and $z$ = 1.5 (bottom panel). CSF protoclusters (red diamonds) show lower values of $\eta_1$ than NR protoclusters (blue triangles) at high redshift.}
\label{deltaz}
\end{figure}

In Fig.\ref{etaeta}  we represent  the relation between  the two definitions of virial ratio, $\eta$ and $\eta_1$,  computed  both for  CSF (top panel) and NR (bottom panel) subsets, for the same redshift bins we previously considered.   
 At $z$ = 4,  the extent of the range   covered by the virial ratio for CSF protoclusters  is much larger than in the NR case,  having  clusters  with values closer to $\eta_1$ = 1, that is expected for  relaxed objects.   On the contrary,  NR clusters at all redshifts   present a behavior of the virial ratios with no differences with redshift,  none of them having  values  smaller  than 1.2, regardless of whether the pressure term is considered or not.  At lower redshifts ($z$ = 1.5)  higher mean values of $\eta$ and $\eta_1$ are found for CSF clusters, confirming the mass dependence of the two parameters, and there are no significant  differences between CSF and NR subsets.  Therefore, we conclude that the effect of the effective pressure term  computed as mentioned above, underestimate the  contribution of bound particles outside our definition of virial radius\footnote{In the AHF halo finder,  the {\em virial }radius of the halos  is defined  as the  radius that encompasses a mean density  that fulfills the  numerical solution of the spherical top hat model at that redshift}. Nevertheless, it is interesting to  see  that there are some  high-z protoclusters in CSF simulations  that do follow  the virial theorem, even  without   correction for pressure terms.

We finally explore the relation between the two criteria adopted here to define the dynamical state of halos, virial ratios ($\eta$,$\eta_1$) and center-of-mass offset ($\Delta r$). In Fig.\ref{deltar} we show the relation between the average values of  $\eta_1$ (and $\eta$) CSF and NR protoclusters at different redshift bins $\Delta r$.  There is a clear relation  between the two dynamical state estimators. The  virial ratios  flatten  off for  values  of $\Delta r \le 0.1$ and the start to increase when   $\Delta r \ge 0.1$.  Therefore this confirms  the validity of both criteria to study the dynamical state of simulated halos.   According to these results, we simply define those protoclusters of our dataset with $\Delta r \leq$ 0.1 as relaxed halos, classifying as disturbed all halos showing an higher value. Therefore,  about 30\%  of protoclusters appear to be disturbed for the whole redshift interval considered. 

As  we  already pointed out  before, it is also clear   from  Fig.\ref{deltar}  that 
CSF protoclusters present  values of $\eta_1$ much closer to 1  than NR protoclusters, while at lower redshift  this difference is much less evident. 
This is better seen in Fig \ref{deltaz} where we compare  the $\eta_1$ as a function of $\Delta r$ for CSF and NR  for high (upper panel) and low redshifts (lower pannel). This  behavior is reflecting the different ways  of mass growth of halos at different redshifts.  At high redshift  
the major accretion of matter to halos is  by  smooth accretion through filaments (\citealt{KER05}, \citealt{MAD08}).
    In the case of  CSF halos, the infalling cooled  gas  is able to  overcome the accretion shocks and  makes it way towards  the center of protoclusters, forming   stars efficiently (\citealt{BIR03}, \citealt{DEK09}),  deepening the potential well. Since most of the accretion is smooth, no major source of  kinetic energy is injected to the protoclusters, and thus, the  matter rapidly virialize in the deeper potentials of CSF clusters as compared with  the shallower potentials of NR clusters.   At later times,   protoclusters grow more due to merging  than  to smooth accretion. Therefore, large quantities of kinetic motions is brought to the protocluster. The result is that the deepening in the potential caused by cooled baryons is not capable of increase the virialization process and the result is that  both CSF and NR   halos get similar values of the virial parameter always higher than 1. The impact of the mass accretion at high redshift was already evident in Fig.\ref{etaM}, where it is shown that for CSF objects $\eta_1$ is more sensitive to the redshift evolution of the total mass.

\section{The extention of the Y-M scaling to protoclusters}
The applicability to protoclusters of scaling relations connecting integrated properties of clusters, such as X-rays luminosity and Sunyaev-Zel'dovich (SZ) effect, has never been investigated. One of the main caveats on this analysis would be the problematics related to the observability of objects at high redshifts (as also discussed in the previous sections of this work).\\
Here we try for the first time to explore the evolution of the $Y-M$ scaling relation at $z >$ 1, with the purpose of checking whether the hypothesis of self-similarity, already well studied for clusters of galaxies, can be applied also to protoclusters.\\
It has been shown that the integrated thermal SZ effect, $Y$, whose definition is recalled here as\footnote{$\Omega$ is the solid angle subtended by the cluster, $D_A$ is the diameter angular distance, $k_b$ is the Boltzmann constant, $\sigma_T$ is the Thomson cross section, $m_e$ is the electron rest mass, $T_e$ is the electronic temperature and $n_e$ the numerical density of electrons.}:
\begin{equation}
Y \equiv \int _{\Omega} yd\Omega=D_A^{-2} \bigg(\frac{k_b\sigma_T}{m_ec^2}\bigg)\int_0^\infty dl \int _An_eT_edA
\end{equation}
 is a robust proxy of the total mass of the cluster, more stable than other proxies in the X-rays band (such as bolometric luminosity and temperature), as it is less affected by the physical processes taking place in the central regions of clusters. Previous works (\citealt{BO07}, \citealt{AGH09}, \citealt{KAY12}, \citealt{IO}) have demonstrated that the $Y-M$ scaling relation, which connects the integrated SZ effect directly to the total mass of the cluster, confirms with good accuracy the hypothesis of self-similarity, showing values extremely close to the self-similar prediction, $A$ = 5/3.

To estimate the integrated $Y$ of our dataset of protoclusters we use the same approach already shown in \cite{IO}, where a detailed analysis of the $Y-M$ scaling relation for massive clusters of galaxies has been performed in the redshift range 0 $\leq z \leq$ 1. As in the case of massive clusters, we build synthetic maps of the Compton $y$-parameter for each protocluster at the different redshifts analyzed and we estimate the $Y$ value integrated inside the virial radius. The choice of the integration up to the virial radius is motivated by the limited angular resolution of the expected observations towards so far objects.

The $Y-M$ scaling relation at a fixed overdensity is studied performing a  best fit of
\begin{equation}
Y_{\Delta} = 10^B\bigg(\frac{M_{\Delta}}{\hMsun}\bigg)^AE(z)^{2/3}[h^{-2}Mpc^{2}]\label{eqYM}
\end{equation}
where $M_{\Delta}$ is the total mass calculated inside the sphere of radius $r_{\Delta}$ that we are considering: in our case $\Delta$=$\Delta_{vir}$, so that $M_{\Delta}$ corresponds to the total virial mass of the protoclusters.
The normalization $B$ is defined as:
\begin{equation}
B = \log \Big[\frac{\sigma _T}{m_ec^2}\frac{\mu}{\mu _e}\bigg(\frac{\sqrt{\Delta_c}GH_0}{4}\bigg)^{2/3}\Big]+ \log f_{g},  \label{B}
\end{equation}
and contains all the constant terms and the gas fraction (where $\mu$ and $\mu_e$ are the mean molecular weights respectively of gas and electrons, see section 4.2 of \citealt{IO} for more details).

\begin{figure}
\centering\includegraphics[angle=0,width=8cm]{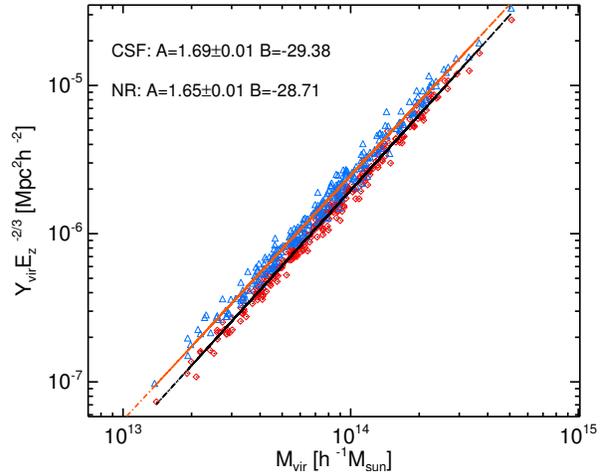}
\caption{$Y-M$ relation for MUSIC protoclusters at $z$ = 1.5: CSF protoclusters are represented by red diamonds (the best fit is the black line) and NR protoclusters by blue triangles (the best fit is the orange line).}
\label{YM15}
\end{figure}

\begin{figure}
\centering\includegraphics[angle=0,width=8cm]{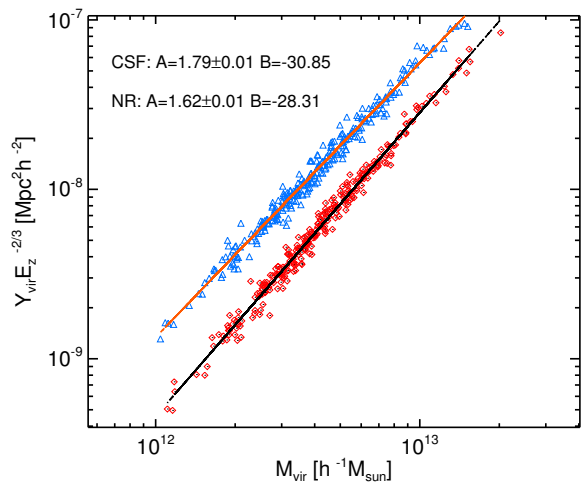}
\caption{$Y-M$ relation for MUSIC protoclusters at $z$=4.0 : CSF objects (red diamonds, best fit is the black line) show a clear deviation from self-similarity, NR objects (blue triangles, best fit is the orange line) have a slope still very close to the self-similar value $A$=5/3. There is also a big difference between the normalizations of the two subsets.}
\label{YM4}
\end{figure}

As in the previous section, in the analysis of the $Y-M$ relation we consider 3 redshifts: 1.5, 2.3 and 4. We find contrasting results. At $z$ = 1.5 (Fig.\ref{YM15}) we find a situation comparable to what already observed at $z\leq$ 1: a slope very close to the self similar value ($A$ = 1.69$\pm$0.01), with no substantial differences between NR and CSF subsets, even if objects simulated with non-radiative physics show higher values of $Y$ and lower slope, as in massive clusters at low redshifts. At $z$ = 2.3 we observe an intermediate situation, with CSF clusters still close to self-similarity ($A$ = 1.70$\pm$0.01) but with a normalization which starts to depart from those of NR objects and of the same objects analyzed at $z <$ 1. At $z$ = 4 (Fig.\ref{YM4}) we find that CSF objects show a much stronger deviation from self-similarity ($A$ = 1.79$\pm$0.01) and values of $Y$ (and of the normalization) much smaller than NR protoclusters, whose scaling relation do not exhibit any 
 significant change from $z$ = 0 even at this high redshift.

Various hypothesis can be made to explain this apparently non self-similar behavior of protoclusters at high redshift. Among these we can remind: the effect of disturbed objects on the scaling relation, an incorrect description of the physical processes taking place in the protoclusters or an effect due to the resolution of the simulation.

\begin{figure}
\centering\includegraphics[angle=0,width=8cm]{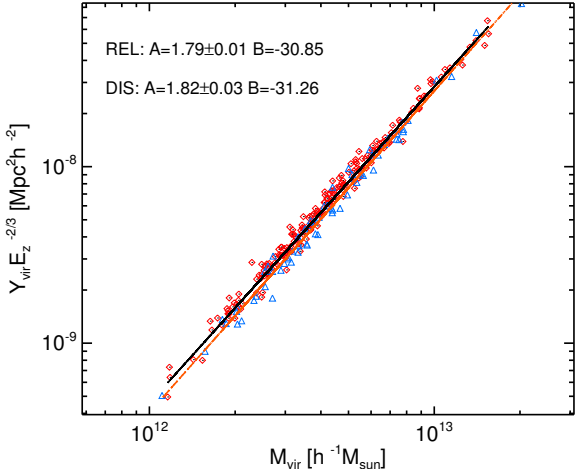}
\caption{$Y-M$ relation for MUSIC protoclusters at $z$ = 4.0 for CSF objects only, distinguished between relaxed (red diamonds, best fit is the black line) and disturbed (blue triangles, best fit is the orange line) objects.}
\label{YM4v}
\end{figure}

Aiming at studying the impact of unrelaxed halos on the $Y-M$ relation, we build two different scaling relations separating relaxed protoclusters from disturbed ones. The results, shown in Fig.\ref{YM4v}, demonstrate that, as it happens for clusters, the dynamical state of the halos does not affect the $Y-M$ scaling relation: both relations exhibit a very similar slope well far from the self-similar value. Moreover, it could be observed that neither the fraction of disturbed objects at high redshift does not differ significantly from the one at low redshifts, nor NR protoclusters analyzed at the same redshifts show any deviation from self-similarity even having the same fraction of disturbed objects (around 30 per cent).

The description of the radiative processes used in the simulation has to be taken into account to check the deviation from self-similarity observed at $z$=4: in fact, the processes taking place in the protoclusters can be different than those used to model clusters at low redshifts. Moreover, MUSIC simulations do not include AGN feedback, which could play a prominent role on gas physics at high redshifts. On the other hand, we have to consider that the effect of AGNs on clusters is usually that of deviating from self-similar conditions and not to get closer to them: therefore it looks quite unlikely that the presence of AGNs could move the scaling relation of protoclusters towards more self-similar values.

The effect of the resolution of the simulation could constitute a non-physical explanation of the deviation from self-similarity: in fact if we consider the mass resolution of MUSIC simulation this allows us to describe massive clusters (with M$_{vir} > 5\times10^{14}$\hMsun) by using several millions of particles. On the contrary, when we move to analyze protoclusters the mass range taken into account is about 3 order of magnitudes smaller (the mean virial mass of our sample at $z$ = 4 is 5$\times10^{12}\hMsun$), resulting into halos described by only a few ten thousands particles, which may be not enough to describe with sufficient precision the integrated properties of protoclusters, such as the integrated $Y$. At the same time, NR protoclusters seem, even if constituted by approximately the same number of particles, not to be affected by the same effects of resolution. 

\begin{figure}
\centering\includegraphics[angle=0,width=8cm]{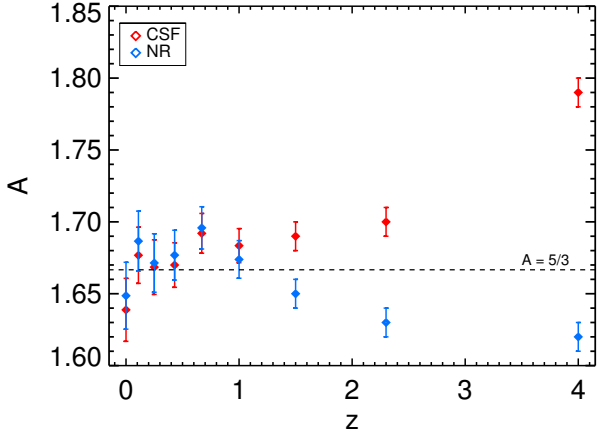}
\caption{Evolution of the slope $A$ of the $Y-M$ scaling relation from $z$=0 to $z$=4: values referring to CSF subset are identified by the red diamonds, the NR subset is represented using blue diamonds.}
\label{A}
\end{figure}

Finally, Fig.\ref{A} shows the evolution of the slope $A$ of the $Y-M$ scaling relation from $z$ = 4 to $z$ = 0, as result of the analysis at high redshifts discussed in this section joined with the analysis performed for massive clusters (whose progenitors are the protoclusters studied in this work) at $z \leq$ 1 by \cite{IO}. We notice how, at the virial radius, clusters keep a very good agreement with self-similarity up to $z$=1.5, starting to depart from it at $z >$ 2, to finally show a clear deviation (in the case of objects simulated including radiative processes) at $z$ = 4. A slight deviation from self-similarity is found also in NR protoclusters.

\section{Summary and Conclusions} \label{sec:conclusions}
The study of the progenitors of clusters of galaxies can give a fundamental contribution to better understand how the massive objects that we observe at present time have evolved. The biggest issue related to the analysis of these objects is related to the difficulties of observing them using present experiments and to the rarity of massive objects at $z >$ 1 predicted by the standard $\Lambda$CDM model. The use of simulations is therefore crucial to limit this problem, making very easy to individuate the high redshift halos which will evolve into clusters.
For the purpose of this work, we adopt the definition of a protocluster as the most massive high redshift progenitor of a galaxy cluster observed at $z$ = 0 and for progenitors as all those high redshift object whose a considerable fraction of mass will be part of the cluster.

Using hydrodynamical simulations of galaxy clusters, here we studied some general properties of protoclusters: the mass and spatial distribution and their redshift evolution, the criteria to distinguish relaxed halos from disturbed ones and the baryon content. We also applied for the first time the study of the $Y-M$ scaling relation to objects at redshifts higher than 1, comparing the results with those referring to clusters at $z \leq$ 1 reported in \cite{IO}.
Our analysis was performed using MUSIC, the largest dataset of hydrodynamically simulated clusters of galaxies at present available. We concentrated on MUSIC-2, an ensemble of 282 lagrangian regions surrounding massive clusters (usually with $M_{vir} >$ 5$\times$10$^{14}$\hMsun) extracted from a big DM only cosmological simulations and resimulated with radiative (CSF subset) and non-radiative (NR subset) physics.
We analyzed protoclusters and progenitors of MUSIC clusters at 3 different redshifts, $z$ = 1.5, 2.3 and 4.0. The main results of our work can be summarized as follows:
\begin{itemize}
\item At $z$ = 4 only a few protoclusters have $M >$ 10$^{13}$\hMsun, while at $z$ = 1.5 we already find more than 100 halos with $M >$ 10$^{14}$\hMsun. At high redshifts, only a fraction (slightly more than 50 per cent at $z$ = 1.5, less then 15 per cent at $z$ = 4) of the mass of the present day cluster is hosted by the progenitors, as most of the mass belongs to diffuse matter or to structure with $M >$ 1.2$\times$10$^{10}$\hMsun. The study of the spatial distribution of protoclusters shows that the cluster forming region, whose center is individuated by the protocluster, has a mean $R_{rms}$ that decreases from about 11$h^{-1}$Mpc at $z$ = 4 to 8 $h^{-1}$Mpc at $z$ = 1.5.
\item The analysis of the baryon content of protoclusters does not show any crucial difference with the results inferred from simulations of galaxy clusters at $z <$ 1. The baryon fraction, normalized to the cosmic ratio and calculated inside the virial radius is $F_b \sim$ 0.95 with no redshift evolution. The normalized gas fraction $F_g$ is ranges from 60 (high redshifts) to 70 per cent (low redshifts) and the star fraction $F_s$ increases with redshift but always with values lower than 40 per cent: the effects of cold flows in MUSIC are therefore limited also at $z >$ 1.
\item We considered different criteria in order to study the dynamical state of protoclusters and to distinguish the relaxed halos from the disturbed ones. Excluding the effect of mergers, that seems to have a smaller impact on high redshift objects, we concentrated on the virial ratio $\eta_1$, corrected including the effect of surface pressure term, and on the spatial shift between the center-of-mass and the center-of-density, $\Delta r$. There is a linear relation between the total mass and the virial ratio, observed only in the CSF subset, and objects at $z$ = 4 show values of $\eta_1$ closer to 1. The two different methods seem to be correlated, as to higher value of $\Delta r$ correspond higher values of $\eta_1$: differently from what observed in DM only simulations, there is a linear dependence between the two parameters. Moreover, the effect of the surface pressure seems to have an impact smaller than in halos simulated only with DM particles. We chose to define as disturbed the ones with $\Delta r \geq$ 0.1.
\item We extended for the first time the analysis of the $Y-M$ scaling relation to objects redshifts higher than 1. While NR protoclusters seem to be in good agreement with the self similar model up to $z$ = 4, on the other hand CSF objects seem to show a deviation from self-similarity at $z >$ 2. The $Y-M$ relation of CSF clusters at $z$ = 4 has a slope $A$ = 1.79, well different from the self-similar expected value $A$ = 5/3 and $Y$ values lower than NR halos. In order to check a possible effect of the dynamical state of objects, we studied the $Y-M$ relation discerning relaxed protoclusters and disturbed ones. No differences have been found between the two subsets. We also made the hypothesis that the deviation from self-similarity may be due to the mass resolution of the simulation, as protoclusters at $z$ = 4 have masses up to three order of magnitudes smaller than clusters at $z$ = 0: anyway NR halos, simulated with the same number of particles, do not show any deviation from self-similarity. Another factor which may contribute to this effect could be an incomplete description of the physical processes taking place inside the protoclusters (i.e. MUSIC simulations do not include AGN feedback): by the way, these factors are expected to have an opposite effect on scaling relations, moving them away from self-similarity.
\end{itemize}

To summarize, the use of hydrodynamical simulations to study protoclusters of galaxies seems very promising to better understand the evolution of present day clusters of galaxies and to approach problematics that are challenging with the resolution of present observational instruments. The proposed large-class satellite mission, PRISM \citep{PRISM}, thanks to the large spectral coverage, angular resolution and sensitivity is expected to deeply explore the universe beyond $z$ =2  planning to detect thousands of objects with $M >$ 5$\times$10$^{13} M_{\odot}$.
The analysis of many interesting protoclusters' properties, such as the dynamical state, the baryon content or the scaling relations, can be considerably improved when using simulations including gas and star particles with respect to DM only simulations.

In order to double check whether the deviation from self-similarity observed in CSF protoclusters at $z$ = 4 is due to real physical effects or it is just a consequence of the resolution of the simulation, we plan to resimulate MUSIC protoclusters in the range 1$\leq z \leq$ 4, improving the mass resolution of at least a factor of 8 and eventually adding more physical processes, such as AGN feedback, and using a binning in redshift narrower than the one adopted in this work.

\section*{Acknowledgements}
The MUSIC simulations were performed at the Barcelona Supercomputing
Center (BSC) and the initial conditions were done at
the Leibniz Rechenzentrum Munich (LRZ).  The authors also thankfully acknowledge the computer resources, technical
expertise and assistance provided by the Red Espa\~nola de Supercomputaci\'on. We thank the support
of the MICINN Consolider-Ingenio 2010 Programme under
grant MultiDark CSD2009-00064. GY acknowledges support
from MICINN under research grants AYA2009-13875-C03-02, AYA2012-31101, 
FPA2009-08958 and Consolider Ingenio SyeC CSD2007-0050.
This work has also been partially supported by funding from the University of Rome Sapienza, Anno: 2012 - prot. C26A12T3AJ.
FS thanks Alexander Knebe for the useful discussions.


\bibliographystyle{mn2e}
\bibliography{archive}

\bsp

\label{lastpage}

\end{document}